\documentclass[12pt]{iopart}
\usepackage{bm}
\usepackage{amssymb}
\usepackage{graphicx}
\usepackage{color}

\newcommand{\pb}{{\bf p}}
\newcommand{\xb}{{\bf x}}
\newcommand{\rb}{{\bf r}}
\newcommand{\Xb}{{\bf X}}

\newcommand{\lal}{{\langle\langle}}
\newcommand{\rar}{{\rangle\rangle}}
\newcommand{\ep}{\varepsilon}

\begin{document}
\title {Quantum fluctuations in modulated nonlinear oscillators}

\author{Vittorio Peano$^1$ and M I Dykman$^2$}
\address{$^1$Institute for Theoretical Physics II, University of Erlangen-Nuremberg,
91058 Erlangen, Germany}
\address{$^2$Department of Physics and Astronomy, Michigan State University, East Lansing, MI 48824, USA}

\begin{abstract}
With a modulated oscillator, we study several effects of quantum fluctuations far from thermal equilibrium. One of them is quantum heating, where quantum fluctuations lead to a finite-width distribution of a resonantly modulated oscillator over its quasienergy (Floquet) states. We also analyze large rare fluctuations responsible for the tail of the quasienergy distribution and switching between the states of forced vibrations. We find an observable characteristic of these fluctuations, the most probable paths followed by the quasienergy in rare events, and in particular in switching.  We also explore the discontinuous change of the most probable switching path where the detailed balance condition is broken. For oscillators modulated by a nonresonant field, we compare different mechanisms of the field-induced cooling and heating of the oscillator.

\end{abstract}

\submitto{\NJP}

\section{Introduction}
\label{sec:Introduction}

The last few years have seen an upsurge in the interest in the dynamics of modulated nonlinear  oscillators \cite{Dykman2012b}. There have emerged several new areas of research where this dynamics plays a central role, such as nanomechanics, cavity optomechanics, and circuit quantum electrodynamics. The vibrational systems of the new generation are mesoscopic. On the one hand, they can be individually accessed, similar to macroscopic systems, and are well-characterized. On the other hand, since they are small, they experience comparatively strong fluctuations of thermal and quantum origin. This makes their dynamics interesting on its own and also enables using modulated oscillators to address a number of fundamental problems of physics far from thermal equilibrium. 

Many nontrivial aspects of the oscillator dynamics are related to the nonlinearity. Essentially  all currently studied mesoscopic vibrational systems display nonlinearity. For weak damping, even small nonlinearity becomes important. It makes the frequencies of transitions between adjacent oscillator energy levels different. Where several levels are occupied, the dynamics strongly depends on the interrelation between the width of the ensued frequency comb and the oscillator decay rate. An important consequence of the nonlinearity is that, when an oscillator is resonantly modulated, it can have coexisting states of forced vibrations, i.e., display bistability \cite{LL_Mechanics2004}. 

One of the general physics problems addressed with modulated nonlinear oscillators is fluctuation-induced switching in systems that lack detailed balance, see \cite{Dykman1979a,Dmitriev1986a,Kautz1988,Vogel1990,Dykman1998,Lapidus1999,Siddiqi2005,Kim2005,Aldridge2005,Stambaugh2006,Almog2007,Mahboob2010} for the classical and \cite{Dykman1988a,Vogel1988,Kinsler1991,Marthaler2006,Peano2006a,Katz2007,Serban2007,Vijay2009,Mallet2009,Peano2010,Wilson2010,Verso2010a} for the quantum regime. A remarkable property of the switching rate in the quantum regime is fragility. The rate $W_{\rm sw}$ calculated for $T=0$, where the system has detailed balance \cite{Drummond1980c}, is exponentially different from the rate calculated for $T>0$, where the detailed balance is broken \cite{Dykman1988a,Marthaler2006}. Recently the effect of fragility of the rates of rare events was also found in the problem of population dynamics \cite{Khasin2009}.  There, too, a small change of the control parameter (infinitesimal, in the semiclassical limit) leads to an exponentially strong rate change.  The nature of the dynamics and the sources of fluctuations in a quantum oscillator and in population dynamics are totally different, and it is important to understand how it happens that they display common singular features.

An important source of quantum fluctuations is the coupling of the oscillator to a thermal bath. It leads to oscillator relaxation via emission of excitations in the bath accompanied by transitions between the oscillator energy levels. The transitions lead to relaxation only on average, in fact they happen at random, giving rise to a peculiar quantum noise. For a resonantly modulated oscillator, the noise causes diffusion over the oscillator quantum states in the external field, which are the quasienergy (Floquet) states. As a result, even where the bath temperature is $T=0$, the distribution over the states has a finite width, the effect of quantum heating \cite{Dykman2011}.

We discuss quantum heating for a resonantly modulated oscillator and compare the predictions with the recent experiment \cite{Ong2013} where the effect was observed. The spectral manifestation of quantum heating is considered, with the focus on the influence of dissipation on the  oscillator spectral characteristic of interest for sideband spectroscopy, the technique which was nicely implemented in  the experiment \cite{Ong2013} using a microwave cavity with an embedded qubit. 

We also study switching between the stable states of forced vibrations of an oscillator modulated close to its eigenfrequency. As quantum heating, switching occurs because of the quantum-noise induced diffusion over the oscillator states. It reminds switching of a classical Brownian particle over the potential barrier due to diffusion over energy \cite{Kramers1940} and therefore is called quantum activation. Generally, the rate of quantum activation largely exceeds the rate of switching via quantum tunneling. We develop an approach to calculating the rate of quantum activation, which naturally connects to the conventional formulation of the rare events theory in chemical and biological reaction systems and in population dynamics \cite{Touchette2009,Kamenev2011}. This approach provides a new insight into the fragility of the switching rate of the oscillator.

The dynamics of a periodically modulated harmonic quantum oscillator coupled to a thermal reservoir is one of exactly solvable problems of physical kinetics \cite{Schwinger1961,Zeldovich1969,Zeldovich1970}. However, the solution disregards the fact that nonresonant modulation can open a new channel for oscillator relaxation, where a transition between the oscillator energy levels is accompanied by an emission (absorption) of an excitation in the medium, while the energy deficit is compensated by the modulation. Alternatively, the role of the medium can be played by another mode with the relaxation rate higher than that of the oscillator  \cite{Dykman1978}. This mechanism underlies the cooling studied in cavity optomechanics. We provide a brief comment in order to unify various mechanisms of the change of the quantum distribution of the oscillator by nonresonant modulation.

\section{Quasienergy spectrum and the master equation}
\label{sec:q_heating}

\subsection{Hamiltonian in the rotating frame}
\label{subsec:model}

A major type of the internal oscillator nonlinearity of interest for the effects we will discuss is the Duffing nonlinearity, where the potential energy has a term quartic in the oscillator displacement $q$; in quantum optics, it corresponds to the Kerr nonlinearity. The simplest types of resonant modulation that lead to the bistability of the oscillator are additive modulation at frequency $\omega_F$ close to the oscillator eigenfrequency $\omega_0$ and parametric modulation  (modulation of the oscillator frequency) at frequency $\approx 2\omega_0$ \cite{LL_Mechanics2004}. The analysis of quantum fluctuations in these two systems has much in common \cite{Dykman2012}, and the method that we will develop here applies to the both types of systems. For concreteness, we will consider here additive modulation. The oscillator Hamiltonian is
\begin{eqnarray}
\label{eq:H_0(t)}
H_0=\frac{1}{2}p^2+\frac{1}{2}\omega_0^2q^2 +\frac{1}{4}\gamma q^4 + H_F(t), \qquad H_F=-qA\cos\omega_Ft,
\end{eqnarray}
where $q$ and $p$ are the oscillator coordinate and momentum, the mass is set equal to one, 
$\gamma$ is the anharmonicity parameter, and $A$ is the modulation amplitude.  We assume that the the modulation is resonant and not too strong, so that
\begin{equation}
\label{eq:delta_omega}
|\delta\omega|\ll \omega_0, \qquad \delta\omega=\omega_{F}-\omega_0; \qquad |\gamma|\langle q^2
\rangle\ll\omega_0^2.
\end{equation}

A periodically modulated oscillator is described by the Floquet, or quasienergy states $\Psi_{\ep}(t)$. They provide a solution of the Schr\"odinger equation $i\hbar\partial_t\Psi=H_0(t)\Psi$ that satisfies the condition $\Psi_{\ep}(t+t_F)=\exp(-i\ep t_F/\hbar)\Psi_{\ep}(t)$, where $t_F=2\pi/\omega_F$. This expression defines quasienergy $\ep$. To find quasienergies and to describe the oscillator dynamics it is convenient to change to the rotating frame. This is done by the standard canonical transformation $U(t)=\exp\left(-ia^{\dag}a\,\omega_Ft\right)$, where $a^{\dag}$ and $a$ are the raising and lowering operators of the oscillator. We introduce slowly varying dimensionless coordinate $Q$ and momentum $P$ in the rotating frame,
%

\[
\fl
U^{\dag}(t)q U(t) = C_{}(Q\cos\varphi_{}+P\sin\varphi_{}), \qquad
U^{\dag}(t)p U(t) = -C_{}\omega_F(Q\sin\varphi_{} - P\cos\varphi_{}).\]
%
Here, $\varphi =\omega_Ft$ and the scaling factor is $C=\left|8\omega_F(\omega_F-\omega_0)/3\gamma\right|^{1/2}$. The commutation relation between $P$ and $Q$ has the form
\begin{equation}
\label{eq:lambda}
[P,Q]=-i\lambda,\qquad \lambda=\hbar/(\omega_F C_{}^2)\equiv 3\hbar|\gamma|/8\omega_F^2|\omega_F-\omega_0| .
\end{equation}
Parameter $\lambda\propto \hbar$ plays the role of the Planck constant in the quantum dynamics in the rotating frame. It is determined by the oscillator nonlinearity, $\lambda\propto \gamma$. For concreteness we assume that $\gamma,\delta\omega >0$; the oscillator displays bistability for $\gamma\delta\omega>0$.

In the range (\ref{eq:delta_omega}) the oscillator dynamics can be studied in the rotating wave approximation (RWA). The RWA Hamiltonian is
\begin{eqnarray}
\label{eq:H_rotating_frame}
\tilde H_0=U^{\dag}H_0 U-i\hbar U^{\dag}\dot{U}\approx
\frac{3}{8}E_{\rm sl}\hat{g}_{}, \nonumber\\
g(Q,P) =
\frac{1}{4}\left(P^2+Q^2 -1\right)^2 - \beta^{1/2}Q,\quad \beta = 3|\gamma|A^2/32\omega_F^3|\omega_F-\omega_0|^3,
\end{eqnarray}
where $\beta$ is the scaled modulation intensity and $E_{\rm sl}=\gamma C_{}^4$ is the characteristic energy of motion in the rotating frame. This motion is slow on the time scale $\omega_F^{-1}$. Note that $E_{\rm sl}\ll \omega_0^2\langle q^2\rangle \sim \omega_0^2C^2$. 

\begin{figure}[h]
\begin{center}
\includegraphics[width=15cm]{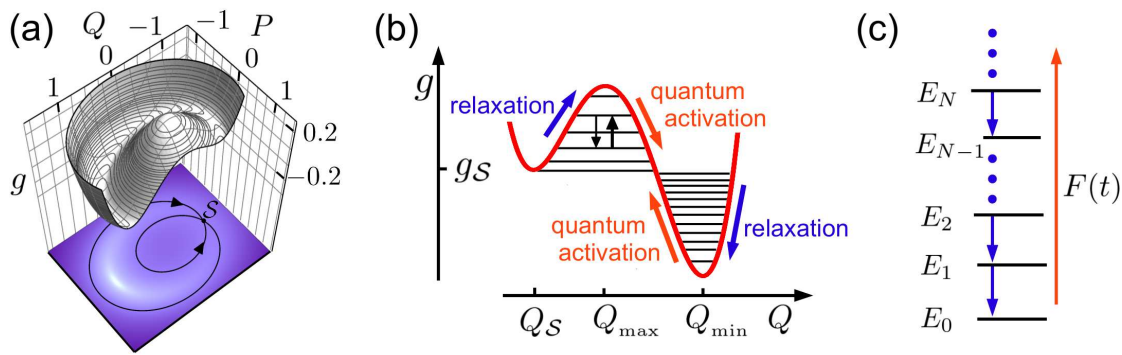} 
\end{center}
\caption{(a) The Hamiltonian function in the rotating frame in the RWA. The extrema of $g(Q,P)$ correspond to the stable vibrational states in the limit of weak damping. (b) The cross-section  $g(Q,0)$ and the quasienergy levels of the states localized about the extrema of $g(Q,P)$. Points $Q_{\min}, Q_{\max}$ and $Q_{\cal S}$ indicate the positions of the minimum, the local maximum, and the saddle point of $g(Q,P)$, respectively. The plots refer to $\beta = 0.01$ and $\lambda = 0.041$.   (c) The transitions between the Fock states of the oscillator with energies $E_N\approx \hbar\omega_0(N+1/2)$ accompanied by emission  of excitations in the bath, e.g., photons. Some of the corresponding transitions between quasienergy states are shown by small arrows in (b).  The stationary state of the oscillator is formed on balance between relaxation and excitation by periodic modulation $F(t)$.  }
\label{fig:quasienergy}
\end{figure}

Operator $\hat g_{}=g(Q,P)$ is the dimensionless Hamiltonian in the rotating frame. In the RWA, the Schr\"odinger equation for the RWA wave function $\psi(Q)$ in dimensionless slow time $\tau$ reads
\begin{equation}
\label{eq:Schrodinger_eq}
i\lambda\dot\psi \equiv i\lambda \partial_{\tau}\psi= \hat g \psi, \qquad \tau= t|\delta\omega|\equiv (\lambda E_{\rm sl}/\hbar)t.
\end{equation}
Operator $\hat g$ is independent of time and has a discrete spectrum, $\hat g|n\rangle=g_n|n\rangle$. The eigenvalues $g_n$ give the quasienergies in the RWA, $\ep_n= (3E_{\rm sl}/8)g_n$ (we are using an extended $\ep$-axis rather than limiting $\ep$ to the analog of the first Brillouin zone $0\leq \ep<\hbar\omega_F$).  Function $g(Q,P)$ has the shape of a tilted Mexican hat and is shown in Fig.~\ref{fig:quasienergy}(a); the quasienergy levels are shown in  Fig.~\ref{fig:quasienergy}(b) .
\noindent

 In contrast to the Hamiltonian $H_0$, $\hat g$ is not a sum of the kinetic and potential energies. As seen from Fig.~\ref{fig:quasienergy}, the eigenstates localized near the local maximum of $g(Q,P)$ correspond to semiclassical orbits on the surface of the ``inner dome" of  $g(Q,P)$; these states become stronger localized as $g_n$ {\em increases} toward the local maximum of $g(Q,P)$. The quasienergy level spacing $\propto \lambda E_{\rm sl}$ is small compared to the distance between the oscillator energy levels in the absence of modulation,
$|\ep_n-\ep_{n+1}|\sim \lambda E_{\rm sl} \ll \hbar\omega_0$.

\subsection{Master equation for linear coupling to the bath}
\label{subsec:master_equation}

The analysis of the oscillator dynamics is often done assuming that the oscillator is coupled to a thermal bath in such a way that the coupling energy is linear in the oscillator coordinate $q$ and thus in the oscillator ladder operators $a, a^{\dag}$ \cite{Schwinger1961}. In this case the coupling energy $H_i$ and the typical relaxation rate $\Gamma$ are of the form
\begin{equation}
\label{eq:linear_relaxation}
H_i= ah_{\rm b}+{\rm H.c.},\qquad \Gamma\equiv\Gamma(\omega_0)=\hbar^{-2}{\rm Re}~\int\nolimits_0^{\infty}dt\langle [h_{\rm b}^{\dag}(t),h_{\rm b}(0)]\rangle_{\rm b}e^{i\omega_0t},
\end{equation}
where $h_{\rm b}$ depends on the bath variables only and $\langle\ldots\rangle_{\rm b}$ denotes thermal averaging over the bath states. Relaxation (\ref{eq:linear_relaxation}) corresponds to transitions between neighboring energy levels of the oscillator in the lab frame, with energy transferred to bath excitations, see Fig.~\ref{fig:quasienergy}(c). The renormalization of the oscillator parameters due to the coupling is assumed to have been incorporated. For a smooth density of states of the bath, resonant modulation does not change the decay rate parameter, $\Gamma(\omega_F)\approx \Gamma(\omega_0)$. However, it excites the oscillator, as sketched in Fig.~\ref{fig:quasienergy}(c). In a stationary state of forced vibrations (in the lab frame) the energy provided by the modulation is balanced by the relaxation. 

To the second order in the interaction (\ref{eq:linear_relaxation}), the master equation for the oscillator density matrix $\rho$ in dimensionless time $\tau$ reads
\begin{eqnarray}
\label{eq:master_eq}
\dot\rho\equiv\partial_{\tau}\rho=&& i\lambda^{-1}[\rho, \hat g]-\hat\kappa\rho, \qquad \hat\kappa\rho= \kappa(\bar n + 1)(a^\dagger a\rho-2a\rho a^\dagger+\rho a^\dagger a)\nonumber\\
&&+\kappa\bar n(a a^\dagger\rho-2a^\dagger\rho a+\rho aa^\dagger), \qquad \kappa = \Gamma/|\omega_F-\omega_0|.
\end{eqnarray}
Here, the term $\propto [\rho,\hat g]$ describes dissipation-free motion, cf. (\ref{eq:Schrodinger_eq}). Operator $\hat \kappa \rho$ describes dissipation and has the same form as in the absence of oscillator modulation \cite{Mandel1995,DK_review84}; $\kappa$ is the dimensionless decay rate, and $\bar n$ is the oscillator Planck number,
\begin{equation}
\label{eq:a_interms_Q}
a=(2\lambda)^{-1/2}(Q+iP),\qquad \bar n \equiv \bar n(\omega_0) =[\exp(\hbar\omega_0/k_BT)-1]^{-1}.
\end{equation}

In the classical limit $\lambda\to 0$ the oscillator described by (\ref{eq:master_eq}) can have one or two stable states of forced vibrations. Their positions in the rotating frame $(Q_{\rm a},P_{\rm a})$ are given by the stable stationary solutions of the classical equations of motion of the oscillator 
\begin{equation}
\label{eq:classical}
\dot Q = \partial_Pg -\kappa Q, \qquad \dot P = -\partial_Q g -\kappa P.
\end{equation}
Equations (\ref{eq:classical}) are, essentially, the mean-field equations for the moments Tr$(Q\rho)$, Tr$(P\rho)$ for $\lambda\to 0$. For small damping $Q_{\rm a}$ and $P_{\rm a}$ are close to the extrema of $g(Q,P)$.

\section{Quantum heating}
\label{sec:quantum_heating}
\subsection{Balance equation}
\label{subsec:balance_equation}

We will concentrate on the oscillator dynamics in the case where the oscillator is strongly underdamped and its motion is semiclassical, 
\begin{equation}
\label{eq:semiclassical}
\lambda \ll 1, \qquad \kappa \ll 1.
\end{equation}
In this case the number of quasienergy states localized about the extrema of $g(Q,P)$  is large, $\propto 1/\lambda$ [the scaled quasienergies of such states $g_n$ lie between the value of $g$ at the corresponding extremum and the saddle point value $g_{\cal S}$ of $g(Q,P)$ in Fig.~\ref{fig:quasienergy}]. Also, the spacing between the levels is large compared to their width, $|g_n-g_{n\pm 1}|\gg \lambda\kappa$. Where the latter condition is met, the off-diagonal matrix elements of the density matrix on the quasienergy states $\rho_{nm}\equiv \langle n|\rho|m\rangle$  ($n\neq m$) are small. To the lowest order in $\kappa$ the oscillator dynamics can be  described by the balance equation for the populations $\rho_{nn}$ of quasienergy states. From (\ref{eq:master_eq})
\begin{equation}
\label{eq:balance_eqn}
\fl
\dot\rho_{nn}=\sum_m\left(W_{mn}\rho_{mm}-W_{nm}\rho_{nn}\right), \qquad W_{mn}=2\kappa\left[(\bar n + 1)|a_{nm}|^2 +\bar n |a_{mn}|^2\right],
\end{equation}
where 
$a_{nm}\equiv \langle n|a|m\rangle$. We disregard tunneling when defining functions $|n\rangle\equiv \psi_n(Q)$, i.e., we use the wave functions localized about the extrema of $g(Q,P)$; the effect of tunneling is exponentially small for $\lambda \ll 1$. We count the localized states off from the corresponding extremum, i.e., for a given extremum the state with $n=0$ has $g_n$ closest to $g(Q,P)$ at the extremum.

An important feature of the rates  of interstate transitions $W_{mn}$ is that, even for $T=0$ (and thus $\bar n = 0$), there are transitions both {\it toward} and {\it away from} the extrema of $g(Q,P)$. This is because the wave functions $|n\rangle$ are linear combinations of the wave functions of the oscillator Fock states, see Fig.~\ref{fig:quasienergy}(c). Therefore, even though relaxation corresponds to transitions down in the oscillator energy in Fig.~\ref{fig:quasienergy}(c), the transitions up and down the quasienergy have nonzero rates. One can show that, for the both extrema of $g(Q,P)$, the rates of transitions toward an extremum  are larger than away from it. Therefore, depending on where the system was prepared initially, it would most likely move to one or the other extremum of $g(Q,P)$. This is why the extrema correspond to the stable states of forced vibrations of the modulated oscillator in the classical limit of a large number of localized states.

For small effective Planck constant $\lambda \ll 1$, the rates $W_{mn}$ can be calculated in an explicit form by finding the matrix elements  $a_{mn}$ in the WKB approximation \cite{Dykman1988a,Guo2013}. The problem is then related to that of classical conservative motion with Hamiltonian $g(Q,P)$ and with equations of motion of the form $\dot Q=\partial_Pg, \dot P=-\partial_Qg$. A significant simplification comes from the fact that the classical trajectories $Q(\tau;g)$ are described by the Jacobi elliptic functions. As a result, $Q(\tau;g)$ is double-periodic on the complex-$\tau$ plane, with real period $\tau_p^{(1)}(g)$ and complex period $\tau_p^{(2)}(g)$. For $|m-n|\ll \lambda^{-1}$ the matrix element $a_{mn}$ is given by the Fourier $m-n$ component  of the function $a(\tau;g_n) = (2\lambda)^{-1/2}[Q(\tau;g_n)+iP(\tau;g_n)]$\cite{LL_QM81}. It can be calculated along an appropriately chosen closed contour on the complex $\tau$-plane and is determined by the pole of $a(\tau; g_n)$. In particular, for the states localized about the local maximum of $g(Q,P)$ we obtain
\begin{equation}
\label{eq:rate_small_attractor}
|a_{n+k\;n}|^2= \frac{k^2\nu_n^4}{2\beta\lambda}
\frac{\exp[ k\nu_n\,{\rm Im}~(2\tau_*-\tau_p^{(2)})]}{|\sinh[ ik\nu_n\,\tau_p^{(2)}/2]|^2}, \quad \nu_n\equiv \nu(g_n)=2\pi/\tau_p^{(1)}(g_n).
\end{equation}
Here, $\tau_*\equiv \tau_*(g_n)$ and $\tau_p^{(2)} \equiv \tau_p^{(2)}(g_n)$  [Im~$\tau_*, {\rm Im}~\tau_p^{(2)}>0$]; $\tau_*(g)$ is the pole of $Q(\tau;g)$  closest to the real axis;
 $\nu(g)$ is the dimensionless frequency of vibrations in the rotating frame with quasienergy $g$. To the leading order in $\lambda$, we have $W_{n\;n+k} = W_{n-k\;n}$ for $n,n\pm k \gg 1$. 

\subsection{Effective temperature of vibrations about a stable state}
\label{subsec:T_e}

Equation~(\ref{eq:rate_small_attractor}) has to be modified for states very close to the extrema of $g(Q,P)$. Near these extrema the classical motion of the oscillator in the rotating frame is harmonic vibrations. One can introduce raising and lowering operators $b$ and $b^\dagger$ for these vibrations (via squeezing transformation) and expand $g(Q,P)$ near an extremum as
\begin{eqnarray}
\label{eq:squeezed_operators}
Q-Q_{\rm a}+iP=(2\lambda)^{1/2}(b\cosh \varphi_* - b^{\dag}\sinh \varphi_*),\nonumber\\
\hat g\approx g(Q_{\rm a},0) + \lambda\nu_0\left(b^{\dag}b+1/2\right){\rm sgn}\partial_Q^2 g,
\qquad \nu_0=\left|\partial_Q^2 g\partial_P^2 g\right|^{1/2}
\end{eqnarray}
[$(Q_{\rm a},P=0)$ is the position of the considered extremum; it is given by equation $\partial_Qg =Q(Q^2-1)-\beta^{1/2}=0$]. The derivatives of $g$ in (\ref{eq:squeezed_operators}) are evaluated at $(Q_{\rm a},P=0)$. Parameter $\phi_*$ is given by equation  $\tanh \varphi_*=(|\partial_Q^2 g|^{1/2}-|\partial_P^2 g|^{1/2})/(|\partial_Q^2 g|^{1/2}+|\partial_P^2 g|^{1/2})$. 

From (\ref{eq:balance_eqn}) and (\ref{eq:squeezed_operators}), near an extremum of $g$ we have 
\begin{eqnarray}
\label{eq:rates_near_extremum}
W_{m+1\;m}=2\kappa(m+1)(\bar n_e+1),\qquad W_{m\;m+1}=2\kappa(m+1)\bar n_e,\nonumber\\
\bar n_e=\bar n\cosh^2\phi_* + (\bar n+1)\sinh^2\phi_*,
\end{eqnarray}
whereas $W_{m\;m+k}=0$ for $|k|>1$. Equation (\ref{eq:rates_near_extremum}) is a familiar expression for the transition rates between the states of a harmonic oscillator coupled to a thermal bath, with $\bar n_e$ being the Planck number of the excitations of this fictitious bath at the frequency of vibrations in the rotating frame $\nu_0\delta\omega$.

From (\ref{eq:rates_near_extremum}), the stationary distribution of the modulated oscillator over its quasienergy states near an extremum of $g(Q,P)$ is of the Boltzmann type, with effective temperature ${\mathcal T}_e=\lambda\nu_0/ \ln[(\bar{n}_e+1)/\bar{n}_e]$ \cite{Marthaler2006,Dykman2012}. In agreement with the qualitative picture discussed above, this temperature is nonzero even where the temperature of the true bath is $T=0$. This is the effect of quantum heating due to quantum fluctuations in a nonequilibrium system. 
\noindent
\begin{figure}[h]
\begin{center}
\includegraphics[width=5cm,angle=0]{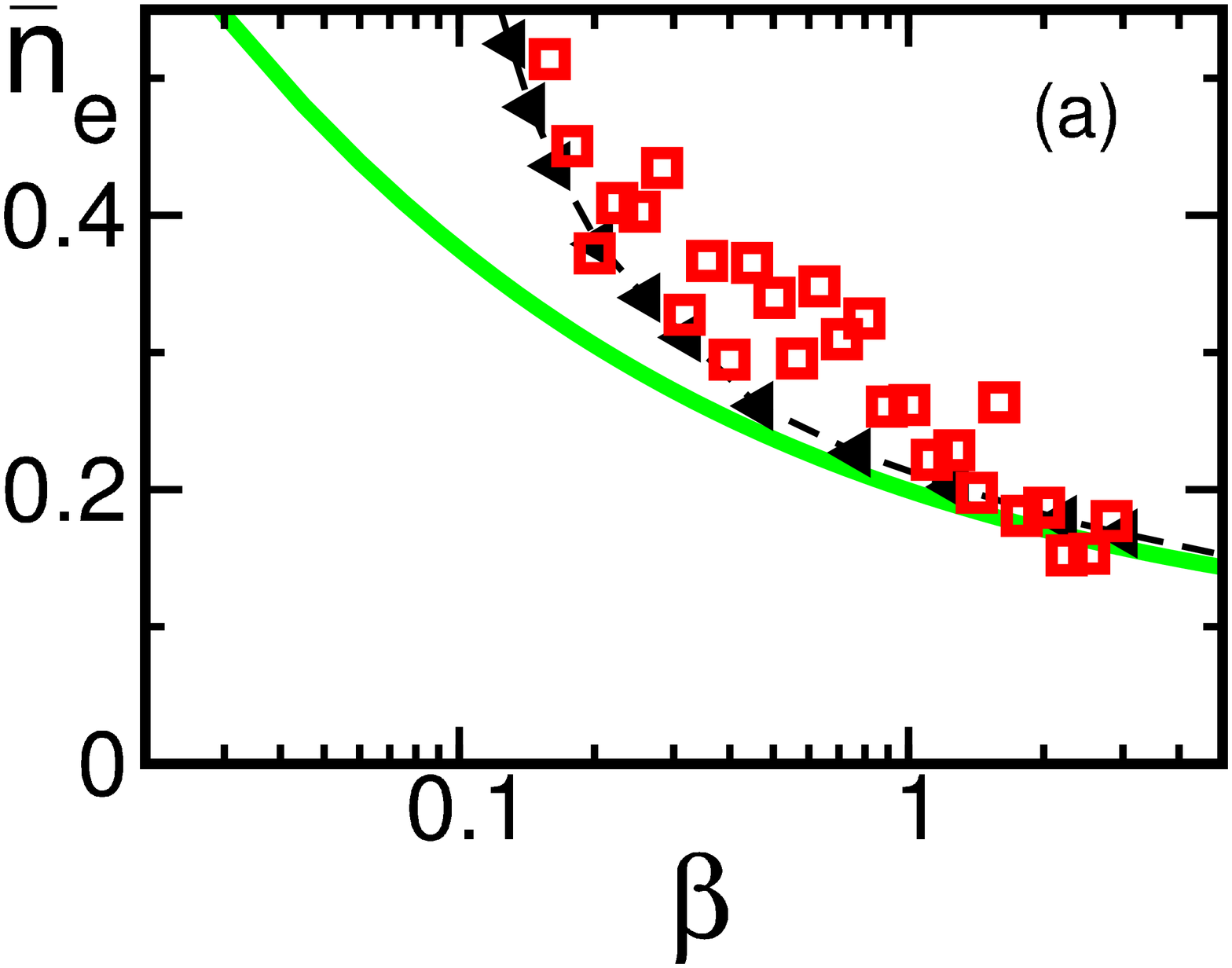} \hspace{2cm}
\includegraphics[width=5.2cm]{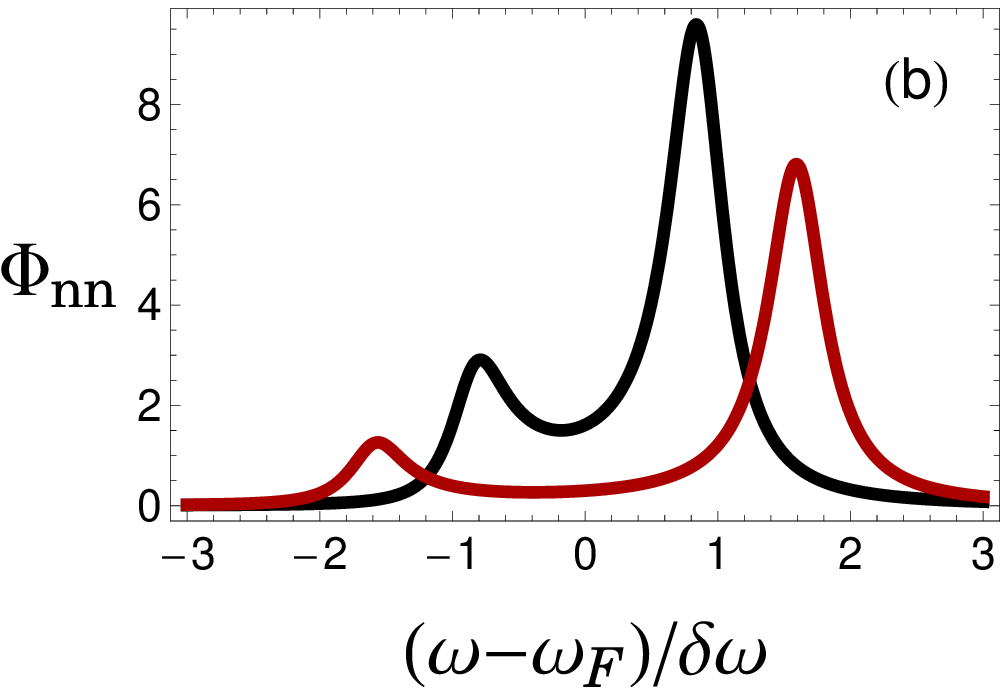}
\end{center}
\caption{(a) The effective Planck number $\bar n_e$  of the vibrations about the large-amplitude state of the modulated oscillator, which corresponds to the minimum of $g(Q,P)$ in the small-damping limit; $\beta$ is the scaled driving field intensity (\ref{eq:H_rotating_frame}). Squares: experimental data \cite{Ong2013}. Solid line:  equation (\protect\ref{eq:rates_near_extremum}) for $\bar n=0$ (see also \cite{Dykman2012}). Triangles: the estimate of the experimentally measured parameter discussed in Appendix A. (b) The scaled power spectra of the oscillator occupation number $\hat n=a^\dagger a$ for vibrations about the large-amplitude stable state for $|\delta\omega|/\kappa = 3.9$ (the value used in \cite{Ong2013}). The black and red curves correspond to $\beta = 0.17$ and 0.8. The triangles in (a) are determined from the ratio $r_\Phi$ of the heights of the lower and higher peaks of $\Phi_{nn}$ as $r_\Phi/(1-r_\Phi)$.}  
\label{fig:q_heating}
\end{figure}

Quantum heating of a resonantly modulated oscillator was recently observed in an elegant experiment \cite{Ong2013} using a mode of a microwave cavity with an embedded Josephson junction \cite{Bertet2012}. The occupation of the excited quasienergy states was revealed using a two-level system (a transmon qubit) as a probe. As seen from Fig.~\ref{fig:q_heating}, the results of the experiment are in a qualitative agreement with the above theory. The agreement improves for larger scaled field intensity $\beta$ (\ref{eq:H_rotating_frame}), where the ratio $\kappa/\nu_0$  is smaller. It is in the range of small $\kappa/\nu_0$ that the quantum temperature is a good characteristic of the distribution over quasienergy states, as the lifetime of these states largely exceeds the reciprocal level spacing (scaled by $\hbar$). Still, even for not too small $\kappa/\nu_0$, the technique developed in \cite{Ong2013} makes it possible to reveal the broadening of the stationary distribution of the modulated oscillator due to quantum fluctuations far from equilibrium. A characteristic of this effect is discussed in Appendix A and the corresponding results are shown in Fig.~\ref{fig:q_heating}.

\section{Switching between the stable states}
\label{sec:switching}

The effect of diffusion over quasienergy states due to quantum fluctuations is not limited to the quantum heating described above. Along with small fluctuations, which lead to comparatively small deviations of quasienergy from its value at an extremum of $g(Q,P)$, there occasionally occur large fluctuations. They push the oscillator far away from the initially occupied extremum. It is clear that, if as a result of such fluctuation, the oscillator goes ``over the quasienergy barrier" to states localized about the other extremum, with probability $\approx 1$ it will then approach this other extremum. Such transition corresponds to switching between the stable states of forced vibrations via the quantum activation mechanism. As seen from Fig.~\ref{fig:quasienergy}(a), with an accuracy to a factor $\sim 1/2$ the switching rate $W_{\rm sw}$ is determined by the probability to reach the saddle-point value $g_{\cal S}$ of $g(Q.P)$. 

The switching rate is small, as switching requires that the oscillator makes many interlevel transitions with rates $W_{mn}$ smaller than the rates of transitions in the opposite direction, $W_{nm}$. Therefore, before the oscillator switches, there is formed a quasistationary distribution over its states localized about the initially occupied extremum of $g(Q,P)$. This is similar to what happens in thermally activated switching over a high barrier \cite{Kramers1940}. However, in contrast to systems in thermal equilibrium, a modulated oscillator generally does not have detailed balance. Its statistical distribution has a simple Boltzmann form with temperature ${\cal T}_e$ only for small damping and only close to the extrema of $g(Q,P)$. Therefore the standard technique developed for finding the switching rate in quantum equilibrium systems \cite{Langer1967,Coleman1977,Affleck1981,Caldeira1983} does not apply. Also, even for $T\to 0$ an oscillator modulated close to its eigenfrequency generally does not switch via tunneling (see \cite{Vogel1988,Peano2006a,Serban2007,Larsen1976,Sazonov1976,Dmitriev1986,Wielinga1993,Marthaler2007a} for the theory of tunneling switching for additive and parametric modulation). Switching via quantum activation is exponentially more probable.

\subsection{Relation to chemical kinetics and population dynamics}
\label{subsec:chem_kinetics}

For small scaled decay rate $\kappa$ the switching rate $W_{\rm sw}$ can be obtained from the balance equation (\ref{eq:balance_eqn}). An approach to solving this equation was discussed earlier \cite{Dykman1988a,Marthaler2006}. Here we provide a formulation that gives an insight into how the oscillator actually moves in switching and also makes a direct connection with the technique developed in chemical kinetics and population dynamics. The balance equation is broadly used in these areas. It describes chemical or biochemical reactions in stirred reactors (no spatial nonuniformity). The reactions can be thought of as resulting from molecular collisions in which molecules change, and if the collision  duration is small compared to the reciprocal collision rate the kinetics is described by a Markov equation \cite{vanKampen_book}
\begin{equation}
\label{eq:chem_kin}
\dot\rho(\Xb,\tau)=\sum_\rb[W(\Xb-\rb,\rb)\rho(\Xb-\rb,\tau)-W(\Xb,\rb)\rho(\Xb,\tau)].
\end{equation}
Here, $\Xb=(X_1,X_2,...)$ is the vector that gives the numbers of molecules $X_i$ of different types $i$, and $\rho$ is the probability for the system to be in a state with given $\Xb$; $W(\Xb,\rb)$ is the rate of a reaction in which the number of molecules changes from $\Xb$ to $\Xb+\rb$. Typically, $X_i$ are large, $X_i\propto N\gg 1$, where $N$ is the total number of molecules. In contrast, the change of the number of molecules in an elementary collision is $|\rb|\sim 1$, because  it is unlikely that many molecules would collide at a time. Equation (\ref{eq:chem_kin}) is also often used in population dynamics, including epidemic models, cf. \cite{ Anderson1992}. In this case the components of $\Xb$ give populations of different species.

Since the number of molecules (population) is large, $N\gg 1$, fluctuations are small on average. Disregarding fluctuations corresponds to the mean-field approximation. In this approximation one can multiply (\ref{eq:chem_kin}) by $\Xb$ and sum over $\Xb$ while assuming that the width of the distribution $\rho(\Xb)$ is small. This gives the equation of motion for the scaled mean number of molecules (population)
\begin{equation}
\label{eq:mean_field_chemical}
\dot{\overline\xb} = \sum_\rb \rb w(\overline\xb,\rb), \qquad \xb = \Xb/N,\qquad w(\xb,\rb)=W(\Xb,\rb)/N.
\end{equation}
Stable solutions of (\ref{eq:mean_field_chemical}) give the stable states of chemical (population) systems. There may be also unstable stationary or periodic states. In population dynamics, an unstable stationary solution of (\ref{eq:mean_field_chemical}) can be the state where one of the species goes extinct.

Equation (\ref{eq:chem_kin}) describes diffusion in the space of variables $\xb$. Along with small $(\propto N^{-1/2}$) fluctuations around the stable states, this diffusion leads to rare large deviations ($\sim {\cal O}(N^{-1})$ in $\xb$-space) and to switching between the stable states. There is an obvious similarity between diffusion over the number of molecules and diffusion over quasienergy states of a modulated oscillator, but there are also some subtle differences, which we discuss below. There is also an obvious difference, with profound consequences: in the case of an oscillator the transition rates $W_{mn}$ (\ref{eq:balance_eqn}), (\ref{eq:rate_small_attractor}) are not limited to $|m-n|\sim 1$.

\subsection{The eikonal approximation}
\label{subsec:eikonal}

The role of the large number of molecules (population) in a modulated oscillator is played by the reciprocal effective Planck constant $\lambda^{-1}$, which determines the number of states localized about the extrema of $g(Q,P)$, cf. Fig.~\ref{fig:quasienergy}. For $\lambda\ll 1$ it is convenient to switch from the state number $n$ to the classical mechanical action $I$ for the Hamiltonian orbits $Q(\tau;g),P(\tau;g)$, which are described by equations $\dot Q=\partial_Pg(Q,P), \dot P=-\partial_Qg(Q,P)$,
\begin{equation}
\label{eq:I_defined}
I=I(g)= (2\pi)^{-1}\int_0^{2\pi/\nu(g)}P(\tau;g)\dot Q(\tau;g)d\tau, \qquad \partial_gI=\nu^{-1}(g),
\end{equation}
where $\nu(g)$ is the vibration frequency for given $g$ [$2\pi I$ gives the area of the cross-section of the surface $g(Q,P)$ in Fig.~\ref{fig:quasienergy}(a) by plane $g=$const]. One can show that, in spite of the nonstandard form of $g(Q,P)$,  the semiclassical quantization condition has the familiar form $I_n\equiv I(g_n)=\lambda (n+1/2)$. 

In the semiclassical approximation the rates of transitions between quasienergy states $W_{mn}$ become functions of the quasicontinuous variable $I$ and can be written as $W_{mn}= W(I_m,n-m)$. The dependence of $W$ on $I$ is smooth, as seen from (\ref{eq:balance_eqn}) and (\ref{eq:rate_small_attractor}), $W(I_m,n-m)\approx W(I_n, n-m)$ for typical $|n-m|\ll 1/\lambda$.

Similar to (\ref{eq:mean_field_chemical}), in the neglect of quantum fluctuations the equation for $\overline I = \sum\nolimits_nI_n\rho_n$ has a simple form
\begin{equation}
\label{eq:mean_field_I}
\dot{\overline I}= \sum_r rw(\overline I,r),\qquad w(I,r)=\lambda W(I,r).
\end{equation}
This equation shows how the oscillator is most likely to evolve. Using that the matrix element $a_{mn}$ in the expression (\ref{eq:balance_eqn}) for the rate $W_{mn}$ is the $(n-m)$th Fourier component of function $(2\lambda)^{-1/2}[Q(\tau;g_m) + iP(\tau;g_m)]$, one can show by invoking the Stokes' theorem that the time evolution of $\overline I$ is extremely simple,
\begin{equation}
\label{eq:deterministic_I}
\dot{\overline I}=-2\kappa \overline I, \qquad \overline I< I_{\cal S},
\end{equation}
where $I_{\cal S}$ is the value of $I(g)$ for $g$ approaching the saddle-point value $g_{\cal S}$ from the side of the extremum of $g(Q,P)$ of interest; the values of $I_{\cal S}$ are different on the opposite sides of $g_{\cal S}$. Equation (\ref{eq:deterministic_I}) coincides with the result for the evolution of $\overline g$ for a classical modulated oscillator \cite{Dykman1979a}. We note that the semiclassical approximation breaks down very close to the saddle point (in particular, the relation $W(I_m,n-m)\approx W(I_n, n-m)$ clearly ceases to apply), but the width of the corresponding range of $I$ goes to zero as $\lambda\to 0$.

We now consider the quasistationary distribution $\rho_n$ about the initially occupied stable state. It is formed on times $(\kappa |\delta\omega|)^{-1}\ll t\ll W_{\rm sw}^{-1}$. To find $\rho_n$ far from the stable state we use the eikonal approximation \cite{Dykman1988a,Marthaler2006}, but in the form similar to that used in chemical kinetics and population dynamics  \cite{Kamenev2011}. We set 
\begin{equation}
\label{eq:eikonal_defined}
\rho_n = \exp[-R(I_n)/\lambda]
\end{equation}
and assume that $|\partial_IR| \ll \lambda^{-1}$. Then $\rho_{n+r}\approx \rho_n\exp[-r\partial_IR]$ for $|r|\ll \lambda^{-1}$ and, to the leading order in $\lambda$, the balance equation (\ref{eq:balance_eqn}) becomes
\begin{equation}
\label{eq:eikonal_Hamiltonian}
\partial_\tau R=-{\cal H}(I,\partial_IR),  \qquad {\cal H}(I,p_I)=\sum_r w (I,r)[\exp(rp_I) -1].
\end{equation}
Equation (\ref{eq:eikonal_Hamiltonian}) has the form of the Hamilton-Jacobi equation for an auxiliary system with coordinate $I$, momentum $p_I$, and action variable $R(I)$ \cite{LL_Mechanics2004}. It thus maps the problem of finding the distribution of the modulated oscillator, which is formed by quantum fluctuations, onto the problem of classical mechanics. The quasistationary distribution is determined by the stationary solution of (\ref{eq:eikonal_Hamiltonian}), i.e., by the solution of equation ${\cal H}(I,\partial_IR)=0$. If there are several solutions, of physical interest is the solution with the minimal $R(I)$, as it gives the leading-order term in $\ln\rho_n$.

\subsection{Optimal switching trajectory}
\label{subsec:instanton}

An advantageous feature of the formulation (\ref{eq:eikonal_Hamiltonian}) is that it provides an insight into how the  quantum oscillator evolves in large fluctuations that lead to occupation of quasienergy states far from the initially occupied extremum of $g(Q,P)$. Even though the diffusion over quasienergy states is a random process and different sequences of interstate transitions can bring the system to the given quasienergy state, the probabilities of such sequences are strongly different. Of physical interest is the most probable sequence, known as the optimal fluctuation. For classical fluctuating systems it has been understood theoretically and shown in experiment and simulations \cite{Kamenev2011,Luchinsky1997b,Hales2000,Ray2006,Chan2008} that the evolution of the system in the optimal fluctuation, i.e., the optimal fluctuational trajectory is given by the classical trajectory of the auxiliary Hamiltonian system, which in the present case is described by equation
\begin{equation}
\label{eq:optimal_trajectory}
\dot I=\partial{\cal H}(I,p_I)/\partial p_I, \; \dot p_I=-\partial {\cal H}(I,p_I)/\partial I;\quad R(I)=\int_0^Ip_IdI. 
\end{equation}
The concept of the optimal fluctuational trajectory  can be extended to the quantum oscillator. Such trajectory for the action variable $I$ is well-defined, since any information of the oscillator phase is automatically erased and the range of the $I$ values largely exceeds the quantum uncertainty in $I$, which is $\propto \lambda$.  Therefore the optimal fluctuational trajectory $I(t)$ can be measured in the experiment in the same way as it is done in classical systems.

In (\ref{eq:optimal_trajectory}) we have set $R(0)=0$ and thus ignored the normalization factor (an analog of the reciprocal partition function) in the expression (\ref{eq:eikonal_defined}) for $\rho_n$. This factor leads to a correction $\propto\lambda$ to $R(0)$. Since, to logarithmic accuracy, the switching rate is determined by the probability of approaching the saddle-point value $I_{\cal S}$ of  $I$, we have
\begin{equation}
\label{eq:R_A_defined}
W_{\rm sw}\sim \kappa |\delta\omega|\exp(-R_A/\lambda), \qquad R_A=R(I_{\cal S}).
\end{equation}
Parameter $R_A$ plays the role of the effective activation energy for switching via quantum activation, with the effective Planck constant $\lambda$ replacing the temperature in the conventional expression for thermally activated switching.

Optimal fluctuations away from the extremum of $g(Q,P)$ are described by optimal trajectories that emanate from $I=0$, which is reflected in (\ref{eq:optimal_trajectory}). The value of the momentum $p_I\equiv \partial_IR$ for $I\to 0$ on the trajectory can be found by noticing that the distribution over quasienergy near the extremum of $g(Q,P)$ is of the form of the Boltzmann distribution with effective temperature ${\cal T}_e$, and thus $R\propto I/{\cal T}_e$; from (\ref{eq:rates_near_extremum}) $p_I= \ln[(\bar n_e +1)/\bar n_e]$ for $I\to 0$. Then from (\ref{eq:eikonal_Hamiltonian}), $\dot I =2\kappa I$ on the optimal trajectory for $I\to 0$. As expected, the system moves along the optimal fluctuational trajectory away from the stable state of fluctuation-free dynamics. 

\begin{figure}[h]
\begin{center}
\includegraphics[width=15cm]{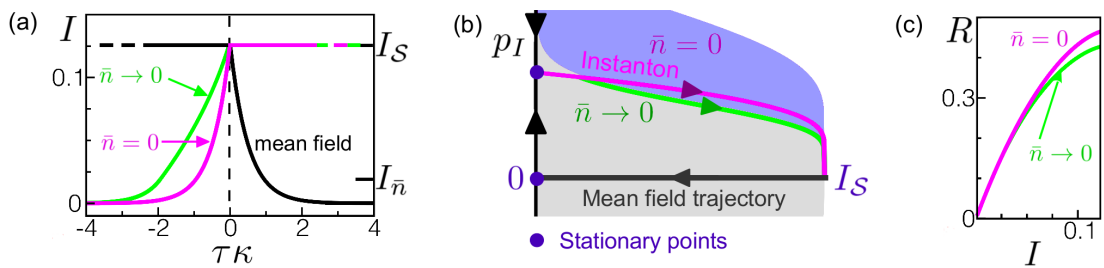} 
\end{center}
\caption{(a) The mean-field (fluctuation free) and optimal fluctuational trajectories of the action variables. Because the system has detailed balance for $\bar n=0$, the optimal trajectory in this case is the time-reversed mean-field trajectory. The data refer to the trajectories for the local maximum of $g(Q,P)$ in Fig.~\protect\ref{fig:quasienergy} for $\beta = 0.035$. The shape of the trajectory changes discontinuously where $\bar n$ becomes nonzero; the trajectories for $\bar n=0$ and $\bar n\to 0$ coincide only for $I<I_{\bar n}$. The limit $\bar n\to 0$ is taken with the constraint $\bar n\gg \lambda^{3/2}$. (b) The phase portrait of the auxiliary Hamiltonian system that describes large fluctuations of the oscillator in the small-damping limit. The real-time instantons in (a) correspond to the trajectories in phase space of the same color. The gray area shows the region where ${\cal H}(I,p_I)$ remains finite for $\bar n\neq 0$; for $\bar n=0$, ${\cal H}$ remains finite in the whole region shown in the figure. (c) The logarithm of the probability distribution $R(I_n)\approx -\lambda\ln\rho_n$ for $\bar n=0$ and $\bar n>0$. }
\label{fig:instanton}
\end{figure}

The facts that $p_I\neq 0$ at the starting point of the optimal trajectory and that the state $I=0$ lies on the boundary of the available values of $I$ are connected with each other and present a distinctive feature of the oscillator dynamics. In chemical kinetics and population dynamics usually stable states lie in the middle of the space of dynamical variables $\Xb$. The probability distribution has a Gaussian maximum at such $\Xb$, and then the momentum on the optimal trajectory is equal to zero \cite{Khasin2009,Kamenev2011}. The states $(I=0,p_I=0)$ and $(I=0,p_I= \ln[(\bar n_e +1)/\bar n_e])$ are stationary points of the Hamiltonian ${\cal H}(I,p_I)$. From (\ref{eq:rates_near_extremum}), $\dot I=\dot p_I= 0$ at these points. The motion of the system near these points is exponential in time and is shown in Fig.~\ref{fig:instanton}.
\noindent

Figure~\ref{fig:instanton}(a) shows the mean-field (fluctuation-free) trajectory $\overline I(t)$ and the optimal trajectory $I(t)$ obtained numerically from equations (\ref{eq:deterministic_I}) and (\ref{eq:optimal_trajectory}), respectively. An interesting feature of the considered model of the modulated quantum oscillator is that it satisfies the detailed balance condition for $T=0$ and thus $\bar n = [\exp(\hbar\omega_0/k_BT)-1]^{-1} =0$ \cite{Drummond1980c}. This is seen from the explicit expression for the rates (\ref{eq:balance_eqn}) and (\ref{eq:rate_small_attractor}), as for $\bar n=0$ they meet the familiar detailed balance condition $W_{n\;n+k}/W_{n+k\;n}=\exp(-k/\xi_n)$ [the explicit form of $\xi_n\equiv \xi(I_n)$ follows from (\ref{eq:rate_small_attractor})]. Therefore $p_I=1/\xi(I)$, and one can show from (\ref{eq:optimal_trajectory}) that $\dot I=2\kappa I$. As a consequence, the optimal fluctuational trajectory $I(t)$ is the time-reversed mean-field trajectory $\overline I(t)$. This is a generic feature of classical systems with detailed balance, see \cite{Luchinsky1997b}. Our results show that the symmetry also holds in quantum systems provided the notion of a trajectory is well-defined.

Of special interest is the vicinity of the  saddle-point value of the action variable $I_{\cal S}$, see Fig.~\ref{fig:instanton}. In a dramatic distinction from chemical kinetics, there is no slowing down of $I(t)$ near $I_{\cal S}$. The quantity $I_{\cal S}$ is a boundary value of $I$ for states localized about a given extremum of $g(Q,P)$ in Fig.~\ref{fig:quasienergy}. Functions $\dot{\overline I}$ and $\dot I$ are discontinuous there. This is an artifact of the balance equation approximation, which applies in the weak damping limit where the dimensionless frequency $\nu(g)\gg \kappa$. For $g\to g_{\cal S}$ the frequency $\nu(g)\to 0$, and the approximation breaks down. With account taken of decay, in the region of bistability the  oscillator has a ``true" unstable stationary state in the neglect of fluctuations. Both the mean-field trajectory and the optimal trajectory in phase space are moving away/approaching this state exponentially in time, cf.~\cite{Dykman1988a}, but the region of $I$ where it happens is very narrow for small $\kappa$.

\subsection{Fragility in the problem of large rare fluctuations}
\label{subsec:fragility}

A striking feature of optimal fluctuational trajectories obvious from Fig.~\ref{fig:instanton} is that these trajectories have different shapes depending on whether the oscillator Planck number is $\bar n=0$ or $\bar n > 0$. The discontinuous with respect to $\bar n$ change of the trajectories and the associated change of the logarithm of the distribution $R(I)$ and of the activation energy for switching $R_A$ show the fragility of the detailed-balance solution for $\bar n =0$ \cite{Dykman1988a,Marthaler2006}. It has been found that the fragility also emerges in a very different type of problem, the problem of population dynamics described by equation (\ref{eq:chem_kin}) \cite{Khasin2009}. In particular, the well-known result for the rate of disease extinction in the presence of detailed balance \cite{Weiss1971,Leigh1981,Jacquez1993,Doering2005a} can change discontinuously with the varying elementary rates $W(\Xb,\rb)$ as the detailed balance is broken.

We now show that the condition for the onset of fragility proposed in \cite{Khasin2009} applies also to the modulated oscillator, even though the divergence it reveals shows up in a different fashion. The condition relies on the expression for the switching exponent. Similar to how it was done for the oscillator, this exponent can be found by seeking the solution of the master equation (\ref{eq:chem_kin}) in the eikonal form $\rho(\Xb)=\exp[-N\tilde R(\xb)]$. To the leading order in $1/N$, the problem is then mapped onto Hamiltonian dynamics of an auxiliary system with mechanical action $\tilde R(\xb)$. From (\ref{eq:chem_kin}), the Hamiltonian of the auxiliary system is
\begin{equation}
\label{eq:chem_Hamiltonian}
{\cal H}(\xb,\pb)=\sum_\rb w(\xb,\rb)[\exp(\rb\pb)-1], \quad\pb = \partial_\xb \tilde R
\end{equation}
[as before, we use that $W(\Xb-\rb,\rb)\approx W(\Xb,\rb)$]. If the system is initially near a stable state $\xb_a$ [a stable solution of (\ref{eq:mean_field_chemical})], $\tilde R(\xb)$ is determined by the Hamiltonian trajectories that emanate from $\xb_a$. From (\ref{eq:chem_Hamiltonian}), $\tilde R(\xb)=\int_{\xb_a}^\xb \pb d\xb$.  The rate of switching from $\xb_a$ (or extinction, in the extinction problem) is $W_{\rm sw}\propto \exp(-N\tilde R_A)$. Similar to the quantum oscillator, 
\begin{equation}
\label{eq:activation_reaction}
\tilde R_A = \int_{\xb_a}^{\xb_{\cal S}}\pb d\xb = \int dt \pb(t)\dot\xb(t)dt.
\end{equation}
Here $\xb_{\cal S}$ is the saddle point of the deterministic dynamics (\ref{eq:mean_field_chemical}); it can be shown that it is the Hamiltonian trajectory  that goes to the saddle point that provides the switching or extinction exponent $\tilde R_A$, cf. \cite{Dykman1979a,Kamenev2011,Herwaarden1995}. Both $\xb_a$ and $\xb_{\cal S}$ are stationary points of the Hamiltonian ${\cal H}$, and the integral over time in (\ref{eq:activation_reaction}) goes from $-\infty$ to $\infty$. This is a significant distinction from the modulated oscillator problem; there equations (\ref{eq:optimal_trajectory}) and (\ref{eq:R_A_defined}) for the activation exponent can be written as 
\[R_A = \int_{-\infty}^0dt p_I(t)\dot I(t),\]
where we set the instant where $I(t)$ reaches $I_{\cal S}$ on the optimal trajectory to be $t=0$.

A small change of the reaction rates $W(\Xb,\rb)\to W(\Xb,\rb)+ \epsilon W^{(1)}(\Xb,\rb)$ ($\epsilon \ll 1$) leads to the  linear in $\epsilon$ change of the Hamiltonian,  ${\cal H} \to {\cal H} + \epsilon {\cal H}^{(1)}$, as seen from (\ref{eq:chem_Hamiltonian}). The action is then also changed. To the first order in $\epsilon$, $\tilde R_A\to \tilde R_A+\epsilon\tilde R_A^{(1)}$. The correction term is given by a simple expression familiar from the Hamiltonian mechanics \cite{LL_Mechanics2004},
\begin{equation}
\label{eq:correction_to_action}
\epsilon\tilde R_A^{(1)} = -\epsilon\int dt {\cal H}^{(1)}\Bigl(\xb(t),\pb(t)\Bigr),
\quad {\cal H}^{(1)}=\sum_\rb w^{(1)}(\xb,\rb)(e^{\pb\rb}-1),
 \end{equation}
where the integral is calculated along the {\it unperturbed} trajectory $\xb(t),\pb(t)$. 
In the extinction problem the integral (\ref{eq:correction_to_action}) can diverge at the upper limit, $t\to\infty$. This is because in this problem $\pb(t)$ remains finite for $t \to\infty$, and therefore if $w^{(1)}(\xb_{\cal S},\rb)$ is nonzero, ${\cal H}^{(1)}\neq 0$ for $t\to\infty$. The divergence indicates the breakdown of the perturbation theory; in the particular example studied in \cite{Khasin2009}, for $\epsilon\to 0$ the change of $\tilde R_A$ was $\sim \tilde R_A$.

For the modulated oscillator, the role of the small parameter $\epsilon$ is played by the Planck number $\bar n$.
If $w^{(0)}(I,r)$ is the transition rate for $\bar n=0$, then from (\ref{eq:balance_eqn}) the thermally induced term in the transition rate has the form $\bar n w^{(1)}(I,r)=\bar n[w^{(0)}(I,r)+w^{(0)}(I,-r)]$. Where the perturbation theory applies, the correction to the effective activation energy of switching reads 
\begin{equation}
\label{eq:correction}
\bar n R_A^{(1)}=-\bar n\int_{-\infty}^0dt\sum_r w^{(1)}\Bigl(I(t),r\Bigr)\{\exp[rp_I(t)]-1\}.
\end{equation}
As we saw, in contrast to reaction/population systems, $p_I\neq 0$ for $t\to -\infty$. However,  $ w^{(1)}(I,r)\propto I\propto \exp(2\kappa t)$ for $t\to -\infty$,  therefore (\ref{eq:correction}) does not diverge for $t\to -\infty$. There is also no accumulation of perturbation for large $t$, as the integral goes to $t=0$. Therefore the cause of the fragility should be different from that in population dynamics/reaction systems.

As mentioned earlier, in contrast to  reaction systems, for the oscillator the values of $r$ in (\ref{eq:correction}) can be large. Then the correction $R_A^{(1)}$ can diverge because of the divergence of the sum over $r$. This happens if on the optimal trajectory $w^{(1)}(I,r)$ decays with $r$ slower than $\exp(rp_I)$. From (\ref{eq:rate_small_attractor}), $w^{(1)}(I,r)$ decays with $r$ exponentially; in particular, $w^{(1)}(I,r)\propto \exp[-2r\nu(g)\tau_*(g)]$ for $r \gg 1$.  The region of the values of $p_I$ where $\sum\nolimits_r w^{(1)}(I,r)\exp(rp_I)$ remains finite is shown in Fig.~\ref{fig:instanton}(b). As seen from this figure, the value of $p_I$ on the $\bar n=0$-trajectory can be too large for the sum over $r$ to converge. Then the perturbation theory becomes inapplicable. The trajectory followed in switching changes discontinuously where $\bar n$ changes from $\bar n=0$ to $\bar n>0$. The probability distribution also changes discontinuously. We note that $|p_I|\sim 1\ll \lambda^{-1}$ on the optimal fluctuational trajectory, which justifies the approximation (\ref{eq:eikonal_Hamiltonian}) that underlies the above analysis. It is clear that the optimal fluctuational trajectory $I(t)$ corresponds to the optimal fluctuational trajectory of the quasienergy $g(t)$, since the $I$ and $g$ variables are related by $\partial_gI=\nu^{-1}(g)$.


The instanton approximation relies on the assumption that the mean square fluctuations provide the smallest scale in the problem, similar to the wavelength in the WKB approximation \cite{LL_QM81}. If the system is perturbed and the perturbation is small, it can be incorporated into the prefactor of the rate of rare large fluctuations. If the perturbation is still small but exceeds the small parameter of the theory, it can be incorporated into the instanton Hamiltonian and leads to a correction to the exponent of the rare event rates. This correction is generically linear in the perturbation. However, this is apparently not a universal behavior, as the unperturbed solution can be fragile with respect to a perturbation. So far the fragility has been found in cases where the perturbation breaks the time-reversal symmetry.

An important problem is the crossover between the instanton solutions without and in the presence of the perturbation. For a modulated quantum oscillator it was recently addressed in \cite{Guo2013} (but the most probable fluctuational trajectories were not studied in this paper). The analysis \cite{Guo2013} shows that the very instanton approximation breaks down by thermal fluctuations, function $\partial_IR$ is not smooth for $\bar n >0$, rather it displays a kink. The threshold for the onset of this behavior is exponentially low in $\bar n$, with $|\ln\bar n|\lesssim \lambda^{-1}$. It corresponds to the regime where the rate of transitions between oscillator states induced by absorption of thermal excitations, which is $\propto \bar n$, becomes comparable with the switching rate $W_{\rm sw}$ calculated for $\bar n=0$. The region where the instanton approximation is inapplicable extends to  $\bar n\lesssim \lambda^{3/2}$.  This is why we indicate that the optimal trajectories in Fig.~\ref{fig:instanton} for $\bar n\to 0$ correspond to vanishingly small $\bar n$ compared to the small parameter of the theory $\lambda$, yet  $\bar n \gtrsim \lambda^{3/2}$.


\section{Nonresonant modulation: a brief summary}
\label{sec:nonresonant}

\begin{figure}[h]
\begin{center}
\includegraphics[width=7cm]{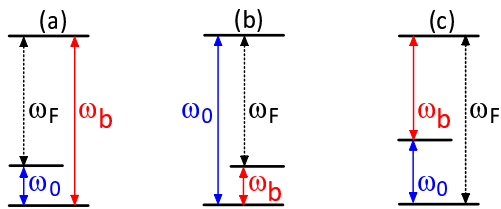} 
\end{center}
\caption{Modulation-induced relaxation processes leading to cooling (a), heating (b), and population inversion (c); $\omega_0$, $\omega_F$, and $\omega_{\rm b}$ are the frequency of the system (the oscillator, in the present case), the modulation frequency, and the frequency of the mode (or a thermal bath excitation) to which the oscillator is coupled by the modulation, respectively; the relaxation time of the mode is much shorter than that of the oscillator. Strong modulation imposes on the oscillator the probability distribution of the fast-decaying mode in (a) and (b) and leads to population inversion in (c). If, in the absence of modulation, oscillator relaxation is described by the standard model (\ref{eq:linear_relaxation}), (\ref{eq:master_eq}), the distribution over the Fock states in the presence of modulation is of the form of the Boltzmann distribution \cite{Dykman1978}; in (c) the distribution over low-energy Fock states is described by negative temperature and the oscillator vibrates close to its eigenfrequency. }
\label{fig:heating}
\end{figure}

Much attention has attracted recently the possibility of cooling mesoscopic oscillators, and the whole new area, the cavity optomechanics, has emerged, see  \cite{Aspelmeyer2013} for a recent review. The cooling is performed by nonresonant modulation, with frequency $\omega_F$ significantly different from the oscillator frequency $\omega_0$. The very idea of cooling different types of quantum systems by a high-frequency field goes back to the mid-70s \cite{Dykman1978,Zeldovich1974,Shapiro1976}, about the same time when the laser cooling of atomic motion was proposed \cite{Hansch1975,Wineland1975}. The change of the distribution can be understood from Fig.~\ref{fig:heating} \cite{DK_review84,Zeldovich1974}. It refers to a system coupled by the modulating field to another system, which can be a thermal bath or a mode with a relaxation time much shorter than that of the system of interest, so that it serves effectively as a narrow-band thermal reservoir.  The modulation provides a new  channel of relaxation for the relatively slowly relaxing system of interest.

Figure~\ref{fig:heating} indicates possible transitions between the states of the system accompanied by energy exchange with the thermal reservoir. For example in (a), a transition of the system from the excited to the ground state is accompanied by a transition of the reservoir to the excited state with energy $\hbar\omega_{\rm b}=\hbar (\omega_0 + \omega_F)$, with the energy deficit compensated by the modulation. On the other hand, a transition of the system from the ground to the excited state requires absorbing an excitation in the thermal reservoir, which is possible only when such excitation is present in the first place. The ratio of the state populations of the system is determined by the ratio of the rates of transitions up and down in energy, and thus by the population of the excited states of the thermal reservoir with energy $\hbar\omega_{\rm b}$.  If the corresponding process is the leading relaxation process, the effective temperature of the system becomes $T^*=(\omega_0/\omega_{\rm b})T$. It means there occurs effective cooling for $\omega_0\ll \omega_{\rm b}$. Similarly, for $\omega_0\gg \omega_{\rm b}$ the modulation leads to heating of the system, see  Fig.~\ref{fig:heating}(b). In the case sketched in Fig.~\ref{fig:heating} (c), the induced transitions from the ground to the excited state of the system are more probable then from the excited to the ground state, which leads to a negative effective temperature for strong modulation.
\noindent
 
In the case of an oscillator, the system has many levels, but the above picture still applies. The goal of this section is to outline and compare different microscopic mechanisms of the coupling of the oscillator to the modulation and the bath.  The unexpected feature is that the distribution of the oscillator over its Fock states can be of the Boltzmann form with an effective temperature determined by the strength and frequency of the modulation \cite{Dykman1978}. However, this is the case only provided the major mechanism of oscillator relaxation in the absence of modulation is the  conventional mechanism (\ref{eq:linear_relaxation}), which in a phenomenological classical description of oscillator dynamics  corresponds to a friction force proportional to the oscillator velocity.

A simple model of the modulation-induced dissipation is where the external field parametrically modulates the coupling of the oscillator to a thermal bath. The coupling Hamiltonian is
\begin{equation}
\label{eq:coupling_parametric}
H_i^{(F)}=-qh_{\rm b}^{(F)}A\cos\omega_Ft.
\end{equation}
Here $h_{\rm b}^{(F)}$ depends on the variables of a thermal bath, or it can be  the coordinate of a comparatively quickly decaying mode coupled to a thermal bath. The interaction (\ref{eq:coupling_parametric}) has the same structure as the interaction (\ref{eq:linear_relaxation}), except that it can lead to decay processes with the energy transfer $\hbar(\omega_0\pm\omega_F)$, cf. Fig.~\ref{fig:heating}(a) and (b). Therefore the structure of the master equation for the oscillator should not change, but the decay parameters and the Planck numbers of excitations created in decay should change appropriately. 
%
The interaction can also lead to decay processes with energy transfer $\omega_F-\omega_0$, for the appropriate modulation frequencies. In this case absorption of bath excitations is accompanied by oscillator transitions down in energy. Respectively, in the master equation (\ref{eq:master_eq}) in the expression for the rates of transitions due to excitation absorption one has to formally replace $\bar n(\omega_0) \to \bar n(\omega_0-\omega_F) = -\bar n(\omega_F-\omega_0)-1$, which means that the friction coefficient becomes negative. 

The above qualitative arguments can be confirmed by a formal analysis similar to that in \cite{Dykman1978}. It shows that in the RWA  the master equation for the oscillator with account taken of the modulation-induced relaxation processes has the form (\ref{eq:master_eq}) with the relaxation parameter $\Gamma$ and the Planck number $\bar n$ replaced by $\Gamma_F=\Gamma + \Gamma_+ + \Gamma_- - \Gamma_{\rm inv}$ and $\bar n_F$, 
\begin{equation}
\label{eq:master_heating}
\partial_t\rho = -\Gamma_F(\bar n_F + 1)(a^\dagger a\rho-2a\rho a^\dagger +\rho a^\dagger a)-\Gamma_F\bar n_F(a a^\dagger \rho-2a^\dagger\rho a + \rho a a^\dagger),
\end{equation}
where
\begin{eqnarray}
\label{eq:Gamma_heating}
\Gamma_{\pm,{\rm inv}}=\frac{A^2}{8\hbar\omega_0}\left|{\rm Re}~\!\int\nolimits_0^{\infty}dt\langle [h_{\rm b}^{(2)}(t),h_{\rm b}^{(2)}(0)]\rangle_{\rm b}e^{i(\omega_0\pm \omega_F)t} \right|,\nonumber\\
\fl
\bar n_F=\left\{\Gamma\bar n(\omega_0)+\Gamma_+\bar n(\omega_0+\omega_F)+\Gamma_-\bar n(\omega_0-\omega_F)+\Gamma_{\rm inv}\left[\bar n(\omega_F-\omega_0)+1\right]\right\}/\Gamma_F.
\end{eqnarray}
Here, $\Gamma_{\pm}$ give the rates of transitions at frequencies $\omega_0 \pm\omega_F$, which correspond to the processes sketched in Fig.~\ref{fig:heating}(a) and (b); $\Gamma_{\rm inv}$ gives the rate of processes sketched in Fig.~\ref{fig:heating}(c), where excitation of the oscillator is accompanied by excitation of the thermal bath. If these processes dominate, they lead to vibrations of the oscillator at frequency $\approx \omega_0$, with amplitude determined by other mechanisms of losses \cite{DK_review84}. Parameters $\Gamma_-$ and $\Gamma_{\rm inv}$ in (\ref{eq:Gamma_heating}) refer to the cases where $\omega_0>\omega_F$ and $\omega_0<\omega_F$, respectively. From (\ref{eq:master_heating}) and (\ref{eq:Gamma_heating}), the probability distribution of the oscillator is characterized by effective temperature $T_F=\hbar\omega_F/k_B\ln[(\bar n_F+1)/\bar n_F]$.

A similar behavior occurs if the modulation is performed by an additive force $A\cos\omega_Ft$, but the interaction with the narrow-band thermal reservoir is nonlinear in the oscillator coordinate, $H_i^{(2)}=q^2h_{\rm b}^{(2)}$. This case was considered in \cite{Dykman1978}. It reduces to the above formulation if one makes a canonical transformation $U(t)=\exp[ v^*(t)a-v(t)a^\dagger ]$ with $v(t)=A_{\rm osc}(2\hbar\omega_0)^{-1/2}(\omega_0\cos\omega_Ft+i\omega_F\sin\omega_Ft)$, where $A_{\rm osc}=A/(\omega_0^2-\omega_F^2)$ is the amplitude of forced vibrations of the oscillator. Indeed, as a result of this transformation $H_i^{(2)}$ transforms into $H_i^{(F)}$ in which the field amplitude $A$ is replaced with $-2A_{\rm osc}$ and $h_{\rm b}^{(F)}$ is replaced with $h_{\rm b}^{(2)}$. 

The analysis of cooling of a vibrating mirror in an optical cavity can be also often mapped onto the analysis for the interaction (\ref{eq:coupling_parametric}). A quantum theory in this case was developed in \cite{Wilson-Rae2007,Marquardt2007}. It considers an oscillator (the mirror) coupled to a cavity mode driven by external radiation. If the radiation is classical, in the appropriately scaled variables the coupling and modulation are described by Hamiltonians $H_i^{\rm (m)}$ and $H_F^{\rm (m)}$, respectively,
\begin{equation}
\label{eq:cavity_mirror}
H_i^{\rm (m)}=c_{\rm b}qq_{\rm b}^2, \qquad H_F^{\rm (m)}=-q_{\rm b}A{}\cos\omega_Ft,
\end{equation}
where $q$ and $q_{\rm b}$ are the coordinates of the mirror and the mode. In cavity optomechanics one usually writes $H_i^{\rm (m)}=c_{\rm b}qa_{\rm b}^{\dag}a_{\rm b}$; the following discussion immediately extends to this form of the interaction.

In the absence of coupling to the mirror, the cavity mode is a linear system, hence $q_{\rm b}(t) = q_{{\rm b}\,0}(t) + [\chi_{\rm b}(\omega_F)\exp(-i\omega_Ft) + {\rm c.c.}]A{}/2$, where $q_{{\rm b}\,0}(t)$ is the mode coordinate in the absence of modulation and  $\chi_{\rm b}(\omega)$ is the susceptibility of the mode \cite{Marquardt2007}. The coupling $H_i^{\rm (m)}$ in the interaction representation then has a cross-term $\propto q_0(t)q_{{\rm b}\,0}(t)\exp(\pm i\omega_Ft)$. Since the cavity mode serves as a thermal bath for the mirror, this term is fully analogous to $H_i^{(F)}$, with $q_{{\rm b}\,0}$ playing the role of $h_{\rm b}^{(F)}$. 

\section{Conclusions}
\label{sec:Conclusions}

It follows from the results of this paper that a modulated nonlinear oscillator displays a number of quantum fluctuation phenomena that have no analog in systems in thermal equilibrium.  Oscillator relaxation is accompanied by a nonequilibrium quantum noise. It leads to a finite-width distribution of the oscillator over its quasienergy states even for the bath temperature $T\to 0$. For resonant modulation, the distribution is Boltzmann-like near the maximum. We have discussed the recent experiment that confirmed this prediction and the effect of oscillator damping on the outcome of a sideband-spectroscopy based measurement of the distribution.

The quantum noise also leads to large rare events that form the far tail of the distribution of the oscillator over quasienergy and to switching between the coexisting states of forced vibrations. We have developed an approach to the analysis of the distribution tail and the switching rate, which makes a direct connection with the analysis of the corresponding problems in chemical and biological systems and in population dynamics. We show that, in a large deviation, the quasienergy of an underdamped oscillator most likely follow a well-defined {\it real} trajectory in {\it real} time. This trajectory is accessible to measurement. For $T=0$, where the oscillator has detailed balance, the most probable fluctuational trajectory is the time-reversed trajectory of the fluctuation-free (mean-field) relaxation of the oscillator to the stable state. Thermal fluctuations break the detailed balance condition and, even where the thermal Planck number $\bar n$ is small compared to the effective Planck constant, lead to an $\bar n$-independent change of the most probable fluctuational trajectory. We show that the criterion of the fragility, i.e., of a discontinuous change of the optimal fluctuation trajectory with the varying parameter can be formulated in a general form, that applies both to reaction systems with classical fluctuations and to the modulated quantum oscillator.

An interesting effect of nonresonant modulation of the oscillator is that it can significantly change the oscillator distribution over the Fock states, leading to heating, cooling, or excitation of self-sustained vibrations depending on the modulation frequency and the coupling to the thermal bath or a mode with a relaxation time shorter than that of the oscillator. We show that different coupling and modulation mechanisms can be described in a similar way and derive explicit expressions for the effective decay rate and temperature of a modulated oscillator.

\ack
We are grateful to P. Bertet and V. N. Smelyanskiy for the discussion. This research was supported in part by the ARO, grant W911NF-12-1-0235, and the Dynamics Enabled Frequency Sources program of DARPA. VP also acknowledges support from the ERC Starting Grant OPTOMECH.

\appendix
\section{Power spectrum of the occupation number of a modulated oscillator}
For some applications, including the experiment \cite{Ong2013}, of interest is the power spectrum $\lal \hat n,\hat n\rar_\omega$ of the occupation number $\hat n = a^\dagger a$ of a modulated oscillator, where
\begin{eqnarray}
\label{eq:power_n_spectrum}
\lal K,L\rar_\omega = \int_0^\infty dt e^{i\omega t}\lal K(t)L(0)\rar\nonumber\\
\fl
\lal K(t)L(0)\rar = \frac{\omega_F}{2\pi}\int_0^{2\pi/\omega_F}dt_i\langle [K(t+t_i)-\langle K(t+t_i)\rangle][L(t_i)-\langle L(t_i)\rangle]\rangle
\end{eqnarray}
(we provide the definition of the relevant correlator for arbitrary operators $K,L$). The major contribution to this power spectrum comes from small-amplitude fluctuations about the stable states of forced vibrations; in the limit of weak damping these states correspond to the extrema of function $g(Q,P)$. We will disregard fluctuation-induced transitions between the stable states and calculate the correlator (\ref{eq:power_n_spectrum}) for each of these states separately; the averaging $\langle...\rangle$ then means averaging over small-amplitude fluctuations about the corresponding state. However, we will not limit ourselves to small damping; moreover, we will assume that the scaled damping rate $\kappa \gg \lambda$, so that the spectra do not display the fine structure related to the nonequidistance of the levels $g_n$ near the stable states \cite{Dykman2011,Dykman2012}. 

Operator $\hat n(t)$ smoothly depends on time: it does not have fast-oscillating factors $\propto\exp(\pm i\omega_0t)$. However, for small $\kappa$,
$\hat n(t)$ contains terms which oscillate at frequencies $\sim\nu_0\delta\omega$ and $\sim 2\nu_0\delta\omega$ with $\nu_0$ being the dimensionless frequency of vibrations about the considered stable state (\ref{eq:squeezed_operators}). To see this, we first note that classical motion about the stable state $(Q_{\rm a},P_{\rm a})$ is described by linearized equations (\ref{eq:classical}) for $\delta Q=Q-Q_{\rm a}, \delta P=P-P_{\rm a}$. For small $\kappa$ this motion is decayed vibrations \cite{Dykman1979a} with dimensionless frequency (the imaginary part of the eigenvalues of the equations for $\delta Q,\delta P$) 
\begin{equation}
\label{eq:nu_a}
\nu_{\rm a}=\left(\kappa^2 + 3r_{\rm a}^4 -4r_{\rm a}^2 +1\right)^{1/2}, \qquad r_{\rm a}^2 = Q_{\rm a}^2 + P_{\rm a}^2.
\end{equation}
From (\ref{eq:squeezed_operators}) and (\ref{eq:nu_a}), $\nu_{\rm a}\to \nu_0$ for $\kappa\to 0$.

We now write the operators of the oscillator as 
\begin{eqnarray}
\label{eq:shifted_a}
a=a_{\rm a}+\delta a, \qquad a_{\rm a}=(2\lambda)^{-1/2}(Q_{\rm a}+iP_{\rm a}),\nonumber\\
\delta a=(2\lambda)^{-1/2}(\delta Q+i\delta P), \qquad \hat n \approx a_{\rm a}^\dagger a_{\rm a} +a_{\rm a}^\dagger \delta a + \delta a^\dagger a_{\rm a} +\delta a^\dagger \delta a.
\end{eqnarray}
This immediately shows that, indeed, $\hat n$ oscillates at dimensional frequencies $\nu_{\rm a}\delta\omega$ and, with smaller amplitude (quadratic in $\delta Q,\delta P$), $2\nu_{\rm a}\delta\omega$. From (\ref{eq:power_n_spectrum}) and (\ref{eq:shifted_a}), the leading term in the power spectrum of $\hat n$ is
\begin{eqnarray}
\label{eq:phi_nn}
{\rm Re}~\lal \hat n,\hat n\rar_\omega \approx (r_{\rm a}^2/2\lambda)|\delta\omega|^{-1}\Phi_{nn}(\omega),\nonumber\\
\Phi_{nn}(\omega)=|\delta\omega|\,{\rm Re}~[\lal \delta a,\delta a^\dagger\rar_\omega + \lal\delta a^\dagger ,\delta a\rar_\omega]
\end{eqnarray}
Functions $\lal \delta a,\delta a^\dagger\rar_\omega$ and $\lal \delta a^\dagger,\delta a\rar_\omega$ were found earlier \cite{Serban2010} (see also \cite{Dykman2012} where the current notations were adopted).  Using their explicit form, one can show that, for small decay rate, $\kappa\ll\nu_{\rm a}$, function $\Phi_{nn}(\omega)$ has two Lorentzian peaks. They are located at $\approx \pm\nu_a|\delta\omega|$ and have halfwidth $\kappa|\delta\omega|$. 

Examples of the spectra $\Phi_{nn}(\omega)$ are shown in Fig.~\ref{fig:q_heating}(b). For the chosen parameters the spectra have two well-resolved peaks. A straightforward but somewhat tedious calculation shows that the ratio of the heights of the peaks of $\Phi_{nn}(\omega)$ approaches $\bar n_e/(\bar n_e+1)$ for $\kappa/\nu_a\to 0$, where $\bar n_e$ is the effective Planck number for vibrations about the stable state (\ref{eq:rates_near_extremum}).  Therefore by measuring this ratio one can reveal and quantitatively characterize the effect of quantum heating. However, for the parameters in Fig.~\ref{fig:q_heating}(b), even though the peaks are well resolved, the ratio of their heights is different from $\bar n_e/(\bar n_e+1)$. This ratio as a function of the scaled intensity of the modulating field $\beta$ is shown by triangles in Fig.~\ref{fig:q_heating}(a).

In the experiment \cite{Ong2013} the occupation of excited quasienergy states of the oscillator was detected by attaching the oscillator to a two-level system (qubit). There was applied an extra field $\propto F_q\cos \omega_qt$ at frequency $\omega_q$ close to the transition frequency of the qubit $\omega_{\rm ge}$, and then the resulting population of the excited qubit state was measured. The frequencies were chosen in such a way that $|\omega_0-\omega_q| \gg |\omega_q-\omega_{\rm ge}|$, therefore transitions between the qubit states accompanied by transitions between the Fock states of the oscillator have negligible probability. However, one can think that the oscillator modulates the coupling of the qubit to the field $F_q$. Then phenomenologically one can write the effective Hamiltonian of the driven qubit as
\begin{equation}
\label{eq:qubit_Hamiltonian}
H_q=\frac{1}{2}\hbar\omega_{\rm ge}\sigma_z -\frac{1}{4} \left[F_q\exp(i\omega_qt)\sigma_- (1+\alpha_q\hat n) + {\rm H.c}\right],
\end{equation}
where $\sigma_z,\sigma_{\pm}=\sigma_x\pm i\sigma_y$ are Pauli matrices. We emphasize that this is a phenomenological ``toy" model in which we keep only resonant terms, $\alpha_q$ is a phenomenological parameter; a discussion of the microscopic model will be performed by the authors of Ref~\cite{Ong2013}, it is beyond the scope of this paper.

Our toy model captures the possibility of transitions between the qubit states induced by the field $\propto F_q$ and accompanied by transitions between the oscillator quasienergy levels. We will assume that the decay rate of the oscillator is larger than the decay rate of the qubit. Then, to the second order in $\alpha_q$, the contribution to the rate of qubit excitation $ |\downarrow\rangle \to |\uparrow\rangle $ due to the qubit-oscillator coupling is given by $(2\hbar)^{-1}|\alpha_qF_q|^2{\rm Re}\lal \hat n,\hat n\rar_\omega$ with $\omega =\omega_q - \omega_{\rm ge}$. The coupling-induced contribution to the rate of qubit transitions  $  |\uparrow\rangle \to |\downarrow\rangle $ is given by the same expression with the correlator evaluated for $\omega = \omega_{\rm ge}-\omega_q$. However,  transitions  $  |\uparrow\rangle \to |\downarrow\rangle $ are more likely to be dominated by spontaneous processes or induced by the $\hat n$-independent term in (\ref{eq:qubit_Hamiltonian}). On the other hand, for $\omega_q-\omega_{\rm ge}$ near the peaks of the correlator $\Phi_{nn}(\omega)$, the coupling-induced transitions $ |\downarrow\rangle \to |\uparrow\rangle $ can have a substantial relative probability and determine the resulting population of the excited qubit state. Then the ratio of the heights of the peaks of $\Phi_{nn}(\omega)$ is given by the ratio of the populations of the excited qubit state for the corresponding frequencies, which was measured in the experiment \cite{Ong2013}.

We note that our toy model leads also to the occurrence of a small peak in the population of the excited state of the qubit for $\omega_q-\omega_{\rm ge}$ close to $2\nu_a|\delta\omega|$, which was reported in \cite{Ong2013}  This peak can be related to the peak in the power spectrum of $\lal \hat n,\hat n\rar_\omega$ for $\omega \sim 2\nu_a|\delta\omega|$, which results from the term $\propto \delta a^\dagger\delta a$ in $\hat n$ in (\ref{eq:shifted_a}); we note that a contribution to this peak comes also from the anharmonicity of the oscillator vibrations about the stable state, i.e., from the higher-order terms in the expansion of $g(Q,P)$ in $Q-Q_{\rm a}, P-P_{\rm a}$ \cite{Andre2012}.

\section*{References}

\providecommand{\newblock}{}

\end{document}